\documentclass[twocolumn,aps,amsfonts,amsmath,prd,nofootinbib]
{revtex4}
 \usepackage[T1]{fontenc}
 \usepackage[latin1]{inputenc}
\usepackage{amssymb}
\usepackage{amsmath}
\usepackage{graphicx,wasysym}
\usepackage[pdftex]{hyperref}

\def\K{K{\"a}hler}
\def\be{\begin{equation}}
\def\ee{\end{equation}}
\def\ba{\begin{array}}
\def\ea{\end{array}}

\makeatletter

\usepackage{epsf}

\def\be{\begin{equation}}
\def\ee{\end{equation}}
\def\ba{\begin{eqnarray}}
\def\ea{\end{eqnarray}}
\def\beas{\begin{eqnarray*}}
\def\eeas{\end{eqnarray*}}
\def\sla{\raise.15ex\hbox{$/$}\kern-.57em}

\begin{document}

\title{\Large\bf New models of chaotic inflation in supergravity}

\author{Renata Kallosh}
\author{Andrei Linde}

\affiliation{Department of Physics, Stanford University, Stanford, CA 94305, USA}


\begin{abstract}
We introduce a new class of models of chaotic inflation inspired by the superconformal approach to supergravity. This class of models allows a functional freedom of choice of the inflaton potential $V = |f(\phi)|^2$. The simplest model of this type has a quadratic potential $m^2\phi^2/2$. Another model describes an inflaton field with the standard symmetry breaking potential $\lambda^2 (\phi^2-v^2)^2$. Depending on the value of $v$ and on initial conditions for inflation, the spectral index $n_s$ may take any value from $0.97$ to $0.93$, and the tensor-to-scalar ratio $r$ may span the interval form $0.3$ to $0.01$. A generalized version of this model has a potential $\lambda^2 (\phi^\alpha-v^\alpha)^2$. At large $\phi$ and $\alpha > 0$, this model describes chaotic inflation with the power law potential $\sim \phi^{2\alpha}$. For $\alpha < 0$, this potential describes chaotic inflation with a potential which becomes flat in the large field limit. We further generalize these models by introducing a nonminimal coupling of the inflaton field to gravity. The mechanism of moduli stabilization used in these models allows to improve and generalize several previously considered models of chaotic inflation in supergravity. 
\end{abstract}

\maketitle

\section{Introduction}

There were many attempts to implement the chaotic inflation scenario \cite{Linde:1983gd} in supergravity. The main problem was related to the \K\, potential ${\cal K}$. The simplest  \K\, potential contained terms proportional to $\Phi\bar\Phi$. The F-term part of the potential is proportional to $e^{\cal K}$, therefore it was typically growing like $e^{|\Phi|^2}$. This was way too steep for chaotic inflation at $\Phi \gg 1$.

One way to overcome this problem is to find flat directions of the inflaton potential in supergravity, see e.g. \cite{Goncharov:1985ka,Gaillard:1995az}. The simplest version of chaotic inflation in supergravity was proposed in Ref. \cite{Kawasaki:2000yn}. The idea was that instead of considering the \K\, potential $\Phi\bar\Phi$ one may consider the  potential $(\Phi+\bar\Phi)^2/2$. This potential, just like the potential $\Phi\bar\Phi$, renders the field $\Phi$ canonically normalized. However, the new \K\, potential has shift symmetry: It does not depend on the field combination $\Phi-\bar\Phi$. Therefore the dangerous term $e^{\cal K}$ also does not depend on $\Phi-\bar\Phi$, which makes the potential flat and suitable for chaotic inflation, with the field $\Phi-\bar\Phi$ playing the role of the inflaton field. The flatness of the potential is broken only by the superpotential $mS\Phi$, where $S$ is an additional scalar field, which disappears on the inflationary trajectory. As a result, the potential in the direction $\Phi-\bar\Phi$ becomes quadratic, as in the simplest version of chaotic inflation.

This model was followed by many related papers on this subject, see e.g. \cite{Kawasaki:2000yn,Yamaguchi:2000vm,Yamaguchi:2001pw,Kawasaki:2001as,Yamaguchi:2003fp,Brax:2005jv,Kallosh:2007ig,Kadota:2007nc,Davis:2008fv,Takahashi:2010ky}. A similar idea was used in the models of chaotic inflation in string theory \cite{Silverstein:2008sg,McAllister:2008hb}.

Recently a new class of inflationary models with a flat direction was proposed in the context of the standard model \cite{Bezrukov:2007ep} and in the NMSSM \cite{Einhorn:2009bh,Ferrara:2010yw,Lee:2010hj,Ferrara:2010in}, with the Higgs field playing the role of the inflaton field. The flatness of the potential in these models appears because of the nonminimal interaction of the inflaton field to gravity \cite{Salopek:1988qh}. 

This class of theories has many interesting and unusual properties, which were recently revealed in the context of the superconformal approach to supergravity \cite{Ferrara:2010yw,Ferrara:2010in}. In particular, there is a very simple class of supergravity theories, canonical superconmformal supergravity (CSS) models. In these models, the kinetic terms in the Jordan frame are canonical and the scalar potential is the same as in global SUSY. Even when the superconformal symmetry of the chiral multiplets is broken, the potential may remain simple, preserving some of the attractive features of the original theory \cite{Ferrara:2010in}.

In this paper we will show that this new class of inflationary models has an interesting relation to the previously considered models with shift symmetry: In the particular case when the inflaton field is minimally coupled to gravity, the \K\, potential in the models considered in \cite{Einhorn:2009bh,Ferrara:2010yw,Lee:2010hj,Ferrara:2010in} does not depend either on $\Phi-\bar\Phi$, or on $\Phi+\bar\Phi$, which explains the flatness of the inflaton potential in this direction. 

The models \cite{Einhorn:2009bh,Ferrara:2010yw,Lee:2010hj,Ferrara:2010in} also require the existence of an additional field $S$. During inflation, this field acquires a tachyonic mass, but it can be stabilized by adding a term $S^4$ to the \K\, potential  \cite{Lee:2010hj,Ferrara:2010in}. In the models considered in \cite{Kawasaki:2000yn,Yamaguchi:2000vm,Yamaguchi:2001pw,Kawasaki:2001as,Yamaguchi:2003fp,Brax:2005jv,Kallosh:2007ig,Kadota:2007nc,Davis:2008fv,Takahashi:2010ky} the field $S$ was not tachyonic, but in many of these models the mass of this field was much smaller than the Hubble constant during inflation. As a result, inflationary fluctuations of these fields could serve as a source of isocurvature perturbations, or nongaussian adiabatic perturbations \cite{Davis:2008fv}. 

 As we will see, by adding the terms 
term $S^4$ to the \K\, potential of these models, just as it was done for the models considered in \cite{Einhorn:2009bh,Ferrara:2010yw,Lee:2010hj,Ferrara:2010in}, one can render the $S$ field heavy, which then reduces the investigation of such models to the investigation of simple one-field models of chaotic inflation.

The superpotential in the models discussed in our paper has a general structure $ W = Sf(\Phi)$, where $f(\Phi)$ is an arbitrary holomorphic function. When the fields $S$ and ${\rm Im}\, \Phi$ are stabilized at their zero values, these models provide {\it a functional freedom of choice of the inflaton potential},
$V(\phi) = |f(\phi/\sqrt 2)|^2$,
where $\phi =\sqrt 2 {\rm Re}\, \Phi$ in a canonically normalized inflaton field. One can also introduce a non-minimal coupling of the inflaton field to gravity, which provides an additional freedom of choice of the inflaton potential in supergravity. 

In this paper, we will discuss the new class of chaotic inflation models in supergravity  and study their observational consequences.

\section{Superconformal  Approach to Supergravity}\label{CSS}
Recently there was a wave of interest in the possibility to implement inflation in the standard model \cite{Bezrukov:2007ep}. The main idea was to introduce a large nonminimal coupling of the Higgs field to the curvature scalar, of the type of $\xi \phi^2 R$. After that, one can make a transformation from the Jordan frame to the Einstein frame, which eliminates the term  $\xi \phi^2 R$ and makes the potential of the Higgs field sufficiently flat to support inflation \cite{Bezrukov:2007ep,Salopek:1988qh}. 

For a long time it was not quite clear whether one can use this mechanism in supersymmetric models. In the standard derivation of the Lagrangian of matter coupled to supergravity, one starts with a superconformal theory and then fixes the conformal compensator to be proportional to the Planck mass. This brings the supergravity directly to the Einstein frame,  after which it is very difficult to recover the original superconformal symmetry.

 In Refs. \cite{Ferrara:2010yw,Ferrara:2010in} a new approach to supergravity was developed. In this approach, one fixes the conformal compensator in a more general and flexible way, which allows to formulate supergravity in the Jordan frame. After applying this procedure, the term in the action containing the curvature scalar $R$ looks as follows: 
 \be
 \sqrt{-g_J} \, {1\over 2}\,  \Omega^2 (z, \bar z) \,  R(g_J) \ .
\label{R} \ee
  Here $g_J$ is metric in the Jordan frame, $\Omega^2 (z, \bar z)$ is the real frame function\footnote{In \cite{Ferrara:2010yw,Ferrara:2010in} we used the notation $\Omega^2 (z, \bar z)= -{1\over 3} \Phi (z, \bar z)$. Here we reserve the letters $\Phi, \bar \Phi$ for the complex scalar field and use $\Omega^2$ for the real frame function.}, which is an arbitrary function of complex scalar fields $z$ and $ \bar z$.  There is an important class of theories where the \K\, potential is
\be  
{\cal K} =-3 \log \Omega^2 (z, \bar z) \, ,
\ee 
or, equivalently  $ \Omega^2 (z, \bar z)= \exp ({-{\cal K}/3})$. The property of these models is that the kinetic term for the scalar fields in the Jordan frame\footnote{There is also an extra kinetic term for scalars from the auxiliary vector fields \cite{Ferrara:2010yw,Ferrara:2010in}, which vanishes when scalars are either real or imaginary.}
 is given by \cite{Ferrara:2010yw,Ferrara:2010in} 
\be
 \sqrt{-g_J} \,  3 \, \Omega^2_{\alpha\bar \beta} {\partial}_\mu z^\alpha {\partial}_\nu \bar z^{\bar\beta } g^{\mu\nu}_J \ ,
\ee
where 
\be
\Omega^2_{\alpha\bar \beta}\equiv {\partial^2 \Omega^2 (z, \bar z)\over \partial z^\alpha \partial \bar z^{\bar \beta}} \ .
\ee 
For the choice 
\be
\Omega^2= 1-{1\over 3}\left(\delta_{\alpha\bar \beta}  z^\alpha \bar z^{\bar \beta}  + J(z) + \bar J(\bar z)\right),
\label{frame}\ee
 where $J(z)$ is a holomorphic function of scalars, one has $3 \, \Omega^2_{\alpha\bar \beta}= - \delta_{\alpha\bar \beta}$, which  means canonical kinetic terms. 

The potential in the superconformal approach to supergravity is also simplified in the Jordan frame as shown in Eq. (3.4) of \cite{Ferrara:2010in}. 
When conformal symmetry is gauge-fixed, the potential coincides with the global SUSY potential for models where the superconformal symmetry of the matter multiplets is preserved ($J(z)=0$) and the F-term potential is $V_{\rm sc} = |\partial W|^2$. 

Even when the superconformal symmetry is broken and $J(z)\neq 0$, the potential in the Jordan frame may  be relatively simple, as one can infer from Eq. (3.4) of \cite{Ferrara:2010in}. For example, suppose that the holomorphic function  $J(z)$ does not depend on a certain direction, let us call it $S$,  and the superpotential is given by 
\be
W= Sf(\Phi)\, , 
\label{cond}\ee
where $f(\Phi)$ is an arbitrary holomorphic function. If  $S$ is stabilized at $S=0$, the potential of the field $\Phi$  in the Jordan frame  is equal to $V_J=|\partial_S W|^2= |f(\Phi)|^2$. In this case  the potential in the Einstein frame at $S=0$ is also simple, namely
\be
V_E|_{S=0}= {V_J\over \Omega^4}= {|f(\Phi)|^2\over (1-{1\over 3}  (\Phi  \bar \Phi  + J(\Phi) + \bar J(\bar \Phi)))^2}\, .
\label{logPot}\ee

A particular example of this class of models is provided by the models of inflation in the NMSSM \cite{Einhorn:2009bh,Ferrara:2010yw,Ferrara:2010in} and by a simplified version of this scenario proposed in \cite{Lee:2010hj}. For the purposes of this paper it will be sufficient to describe the simple version of these models containing two scalar fields, $S$ and $\Phi$. In these models  the holomorphic $S$-independent function $J(z)$ in $\Omega^2$ in (\ref{frame})  is given by $-{3\chi\over 4 }\Phi^2$. This yields
\begin{equation}
\Omega^2 (\Phi, S; \bar \Phi, \bar S) = 1 -{1\over 3} (\Phi\bar\Phi +  S \bar S)+ {\chi\over 4 }(\Phi^2+ \bar\Phi^2)  + ... ,
\label{K0}\end{equation}
which corresponds to the  \K\, potential 
\begin{equation}
\mathcal{K} = -3\log \left[ 1 - \frac{1}{3}(\Phi \bar\Phi + S \bar S)+ {\chi\over 4 }(\Phi^2+ \bar\Phi^2)+ ...\right] .
\label{K1}
\end{equation}
For the frame function (\ref{K0}),  kinetic terms in the Jordan frame are canonical,
\be
 - \sqrt{-g_J}({\partial}_\mu\Phi {\partial}_\nu \bar \Phi +{\partial}_\mu S {\partial}_\nu \bar S) g^{\mu\nu}_J \ .
\ee
The main goal of the models \cite{Einhorn:2009bh,Ferrara:2010yw,Lee:2010hj,Ferrara:2010in} was to consider strongly nonminimal coupling of the inflaton field to gravity, $\chi \gg 1$, which allows to obtain flat potential at large values of the Higgs field in the NMSSM \cite{Salopek:1988qh,Bezrukov:2007ep}. In the next section we will consider the model of the type considered in  \cite{Einhorn:2009bh,Ferrara:2010yw,Lee:2010hj,Ferrara:2010in}, but we will no longer assume that the field $\Phi$ is the Higgs field of the standard model, and instead of the model with $\chi \gg 1$ we will consider the inflaton field minimally coupled to gravity. 

In order to do it, we will represent Eqs.  (\ref{K0}) , (\ref{K1}) in the following equivalent way \cite{Ferrara:2010in}:
\ba
\Omega^2 &=&  \Bigr[
1   -{1\over 3}S \bar S + {1\over 12}\left (1+  {3\over  2}\chi \right)  (\Phi-\bar \Phi)^2 \nonumber\\ &-& {1\over 12}\left (1-  {3\over 2}\chi \right)  (\Phi+\bar \Phi)^2 +...\Bigl],
\label{AL1} \ea
\ba
\mathcal{K} &=& -3\log \Bigr[
1   -{1\over 3} S \bar S + {1\over 12}\left (1+  {3\over  2}\chi \right)  (\Phi-\bar \Phi)^2 \nonumber\\ &-& {1\over 12}\left (1-  {3\over 2}\chi \right)  (\Phi+\bar \Phi)^2 +...\Bigl].
\label{AL} \ea
One can easily see that this class of models allows some interesting special regimes. For  $\chi = 0$, all scalar fields in the Jordan frame described by Eq. (\ref{K0}) are conformally coupled to gravity. Meanwhile for $\chi = + 2/3$ the real part of the field $\Phi$, which is given by $(\Phi+\bar \Phi)/2$, is not coupled to the curvature scalar $R$, i.e. it is minimally coupled to gravity. In this case the \K\, potential also does not depend on $\Phi+\bar \Phi$. Similarly, for $\chi = - 2/3$ the imaginary part of the field $\Phi$, which is given by $(\Phi-\bar \Phi)/2i$, is minimally coupled to gravity and the \K\, potential does not depend on $\Phi-\bar \Phi$. In this paper we will concentrate on the models with $\chi = + 2/3$, in which case the \K\, potential does not depend on $\Phi+\bar \Phi$; a generalization to the case when it does not depend on $\Phi-\bar \Phi$ is trivial.

The complex fields $S$ and $\Phi$ can be represented as
\begin{equation}
S=(s+i\alpha)/\sqrt 2\,,\qquad \Phi=(\phi+i\beta)/\sqrt 2\,.
\end{equation}
For $\chi = + 2/3$, the field $\phi$ will play the role of the inflaton field. In this case, the \K\, potential, as well as $\Omega$, does not depend on $\phi$. Thus, $\phi$ is minimally coupled to gravity; the Jordan frame for the field $\phi$ coincides with the Einstein frame. If during inflation we will have $S =0$, $\beta = 0$, then the \K\, potential vanishes along the inflationary trajectory, and all fields have canonical kinetic terms. An expression for the potential of the inflaton field along the inflationary trajectory for $W= Sf(\Phi)$ is 
\be
V(\phi) = |f(\phi/\sqrt 2)|^2 \ .
\label{global}\ee
It follows from (\ref{logPot}) with  $\Omega^2=1$ at the inflationary trajectory, when $V_J=V_E$. Depending on the choice of the function $f$, one can obtain a wide variety of models of chaotic inflation in supergravity. 

In the models with $\chi \not = {2/3}$, the frame function $\Omega$ contains a term $(\chi -2/3) (\Phi +\bar\Phi)^2/4$. This term introduces a nonminimal coupling for the inflaton field, ${\xi\over 2} \phi^2 R$, where 
\be
\xi = -{1\over 6} +  {\chi\over 4}.
\ee 
In this case, the potential $V(\phi)$ for $S =0$, $\beta = 0$ in the Jordan frame is still given by (\ref{global}), but it looks different in the Einstein frame,
\be\label{nc}
V_E(\phi) = {|f(\phi/\sqrt 2)|^2\over (1+\xi \phi^2)^2} \ .
\ee
This expression for the Einstein frame potential follows from (\ref{logPot}) with  $\Omega^4=(1+\xi \phi^2)^2$ at the inflationary trajectory, where $V_E= V_J/\Omega^4$. One can also
 obtain this result  by using the standard supergravity expression for $V_E(\phi)$.
It is this modification of the potential that may allow inflation in the standard model along the lines of \cite{Bezrukov:2007ep,Salopek:1988qh,Einhorn:2009bh,Ferrara:2010yw,Lee:2010hj,Ferrara:2010in}. 

Note, that the simplicity of the transition from the Jordan frame potential $|f(\phi/\sqrt 2)|^2$ to the Einstein frame (\ref{nc}) is somewhat deceptive: The field $\phi$ in the Einstein frame with $\xi \not = 0$ is not canonically normalized. The canonically normalized inflaton field $\varphi$ is related to the field $\phi$ as follows:
\be\label{canon}
{d\varphi\over d\phi}  = {\sqrt{1+ \xi\phi^2 + 6\xi^2\phi^2}\over 1+\xi \phi^2} \ .
\ee

It is instructive to compare this new class of models with the previously considered  models of chaotic inflation in supergravity, with the power-law  \K\, potential \cite{Kawasaki:2000yn,Yamaguchi:2000vm,Yamaguchi:2001pw,Kawasaki:2001as,Yamaguchi:2003fp,Brax:2005jv,Kallosh:2007ig,Kadota:2007nc,Davis:2008fv,Takahashi:2010ky},
\be
{\cal K} = \delta_{\alpha\bar \beta}  z^\alpha \bar z^{\bar \beta}  + J(z) + \bar J(\bar z)
  + ... 
\ee
We now make the same assumptions as before, namely that $\partial_S J = 0$, $W= Sf(\Phi)$, $S$ is stabilized at $S = 0$.
In this case  
\be
V_E|_{S=0}= e^{\cal K} |\partial_SW|^2 \ .
\ee
In particular, for
\be
{\cal K} = (\Phi\bar\Phi +  S \bar S)-  {3 \chi\over 4 }(\Phi^2+ \bar\Phi^2)  + ... 
\ee
one has
\be\label{exp}
V_E|_{S=0}=e^{\Phi\bar\Phi -  {3 \chi\over 4 }(\Phi^2+ \bar\Phi^2)}  |f(\Phi)|^2\, .
\ee
For $\chi  = {2/3}$ (i.e. for $\xi = 0$), there is a shift symmetry: $\mathcal{K} =  S \bar S - \frac{1}{2}(\Phi -\bar\Phi)^2 +...$, see \cite{Kawasaki:2000yn}. During inflation with $S = 0$ and $\Phi =\bar\Phi$, one has  $\mathcal{K} =0$, and the potential $V(\phi)$ is the same as in the logarithmic case (\ref{global}). However, in this class of models the general case $\chi  \neq  {2/3}$ does not have a simple interpretation in terms of a nonminimal coupling to gravity.  If we take $\chi  \neq  {2/3}$  in the models with the power-law  \K\, potential, the inflaton potential becomes
\be\label{ncpol}
V(\phi) = e^{-3\xi\phi^2} |f(\phi/\sqrt 2)|^2 \ ,
\ee
where, as before, $\xi =  -{1/6} +  {\chi/4}$. Depending on the sign of $\xi$, the potential either vanishes or blows up at large $\phi$. Therefore it is difficult to achieve inflation in the models with the power-law \K\, potentials with $\xi  \neq  0$. 

In what follows, we will analyze several different models of chaotic inflation with the logarithmic \K\, potentials, as well as their counterparts with the power-law \K\, potentials.

\section{Chaotic inflation with the logarithmic  \K\, potential}\label{log}

We will begin with a model with the superpotential
\begin{equation}
W=-\lambda S(\Phi^2 -v^2/2).
\label{W}\end{equation}
For $v\ll 1$, this model coincides with the model considered in \cite{Einhorn:2009bh,Ferrara:2010yw,Lee:2010hj,Ferrara:2010in}, but now we will consider a more general case and allow $v$ to take any values.
The  \K\, potential is
\begin{equation}
\mathcal{K} = -3\log \left[ 1 + \frac{1}{6}(\Phi -\bar\Phi)^2 - \frac{1}{3}S \bar S+ \zeta (S\bar S)^2/3\right] .
\label{K}
\end{equation}
This corresponds to the choice $\chi = 2/3$ in (\ref{AL}), i.e. to $\xi = 0$.

The scalar potential in this theory depends on $S=(s+i\alpha)/\sqrt 2$ and $\Phi=(\phi+i\beta)/\sqrt 2$.
The field $\phi$ will play the role of the inflaton field. The \K\, potential does not depend on it. The field $S$ plays an auxiliary role. It provides the required potential to the field $\Phi$, but during inflation $S = 0$. An additional term $\zeta (S\bar S)^2$ is introduced to ensure stability of the inflationary trajectory near $S = 0$ \cite{Lee:2010hj,Ferrara:2010in}.   Once stabilization is achieved, the field $S$ vanishes, and this term becomes irrelevant for the  investigation of the dynamical evolution of the field $\phi$.

In this simple model, the potential is a function of $|S|^2$, so it is sufficient to study the potential as a function of $\phi,\, \beta$ and $|S|$ at $S = 0$. An amazing property of this theory, explained in \cite{Ferrara:2010yw,Ferrara:2010in}, is that the kinetic terms of the fields $\Phi$ and $S$ are canonical in the vicinity of the inflationary trajectory $S= 0$, which enormously simplifies investigation of the inflationary regime. 

\begin{figure}[ht!]
\centering
\includegraphics[scale=0.13]{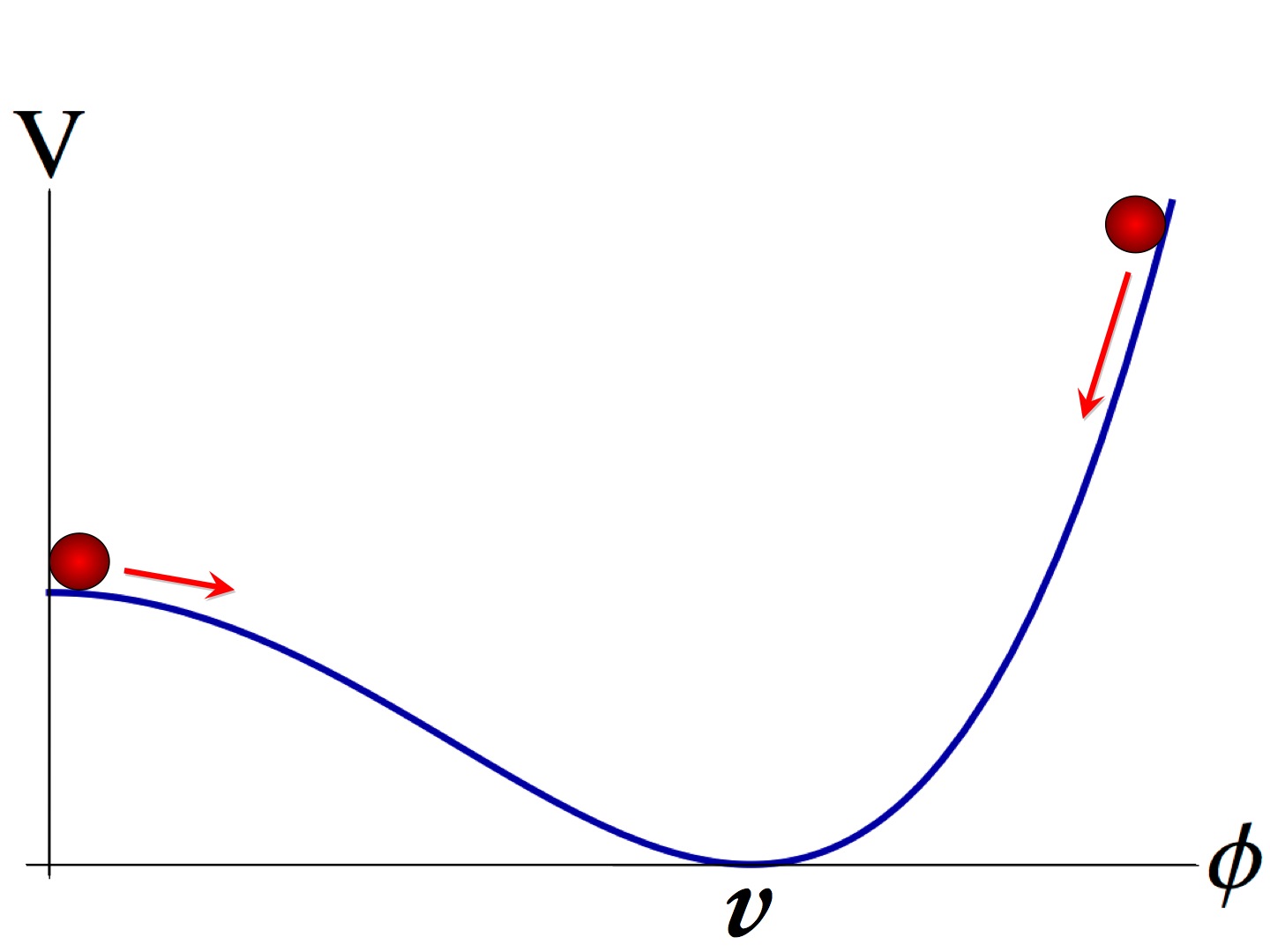}
\caption{Scalar potential for for $s,\,\gamma = 0$ is given by $V(\phi)= {\lambda^2} (\phi^2-v^2)^2/4$. Inflation occurs either when the field $\phi$ rolls down from its large values, as in the simplest models of chaotic inflation, or when it rolls down from $\phi = 0$, as in new inflation.}
\label{potential}
\end{figure}

\begin{figure}[t]
\centering
\includegraphics[scale=0.17]{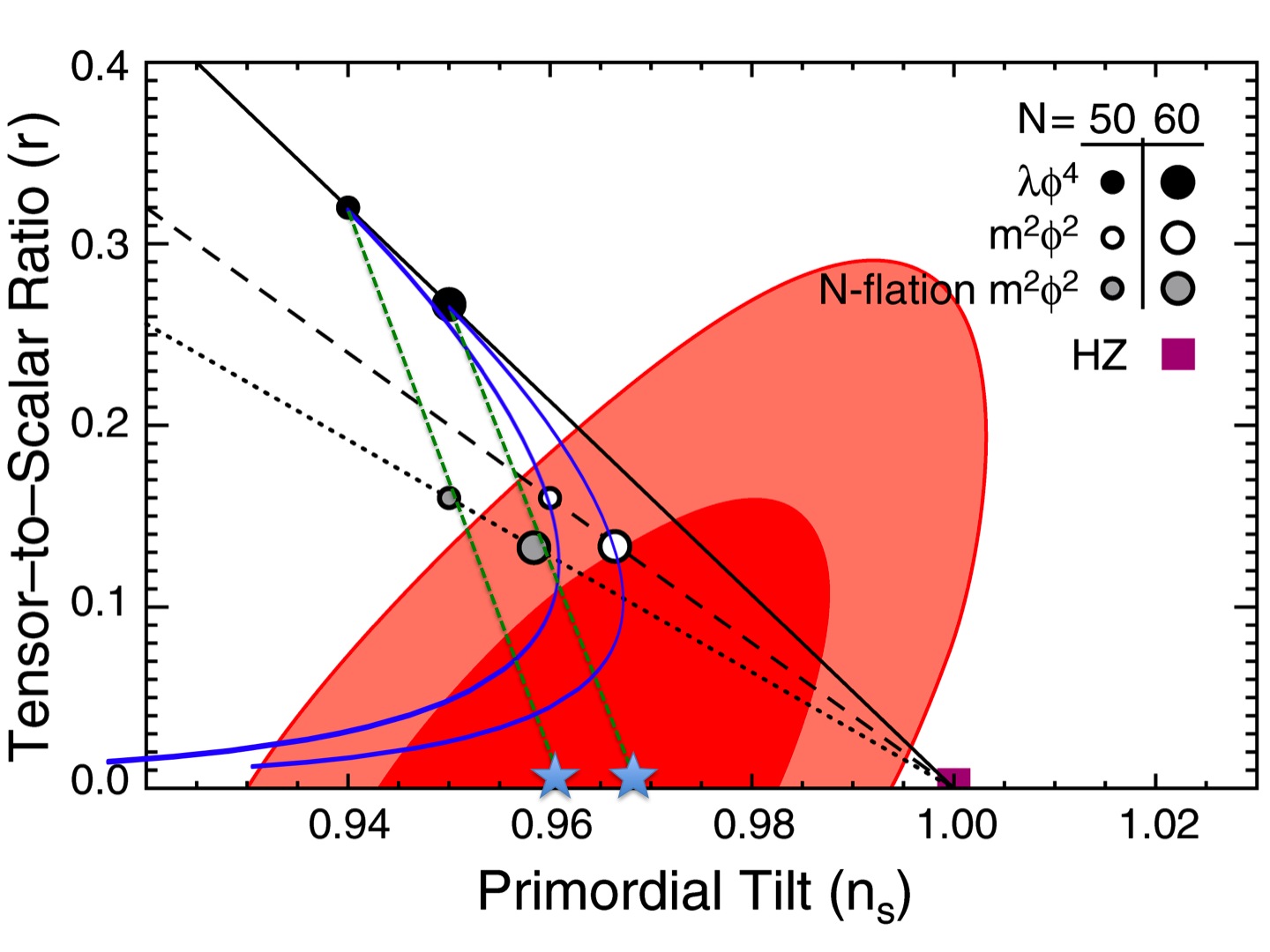}
\caption{WMAP and predictions of our models. Predictions of the model with the potential ${\lambda^2\over 4} (\phi^2-v^2)^2  $ (\ref{inflpot}) are bounded by the two blues lines corresponding to the number of e-foldings $N = 50$ and $N = 60$ \cite{Kallosh:2007wm}. The blue stars correspond to this model with $v = 0$ and the inflaton nonminimally coupled to gravity with $\xi \gg 1$ for $N = 50$ and $N = 60$ \cite{Bezrukov:2007ep,Einhorn:2009bh,Ferrara:2010yw,Lee:2010hj,Ferrara:2010in}. The green dashed lines describe predictions of this model for $v = 0$ and for various values of $\xi$ \cite{Okada:2010jf}, for $N = 50$ and $N = 60$. These results suggest that once one considers arbitrary values of $v$ and increases $\xi$, the blue lines will move to the right and down, as indicated by the green dashed lines. As a result, possible values of $n_s$ and $r$ in this model may span a substantial part of the allowed values of $n_s$ and $r$.}
\label{WMAP}
\end{figure}

As we will see shortly, under certain conditions, the inflationary trajectory corresponds to $S=\beta = 0$. In this case the potential with respect to the scalar field $\phi$ is given by the simple and familiar expression describing the theory with spontaneous symmetry breaking, 
\begin{equation}\label{inflpot}
V(\phi) = {\lambda^2\over 4} (\phi^2-v^2)^2 .
\end{equation}
After inflation the field $\phi$ falls to the minimum of its potential at $\phi = v$, where $V(\phi)$ vanishes. Note that the potential in our model is proportional to $\lambda^2$ rather than to $\lambda$.  Inflation may occur at $\phi \gg 1$, as in the simplest versions of the chaotic inflation scenario  \cite{Linde:1983gd}.     For $v < 1$, $\phi \gg 1$, the potential is $V \approx {\lambda^2\over 4} \phi^4$, so it is ruled out by the WMAP data, see Fig. \ref{WMAP}. However, for $v \gg 1$ during the last stages of inflation the field moves in a quadratic potential near the minimum, with the effective mass $m^2 = 2\lambda^2 v^2$. In this case, COBE normalization requires $m = \sqrt 2 \lambda v \sim 2\times 10^{-6}$ \cite{book,Linde:2007fr}. For $v \gg 1$, inflation may also occur while the field slowly rolls down from the local maximum of its potential at $\phi = 0$, as in the new inflation scenario \cite{new}.

To check whether an inflationary trajectory with $\beta =0$ at $S=0$ is stable with respect to fluctuations of the fields  $\beta$ at $S$, let us first study the potential of the fields $\phi$ and $\beta$ at $S =0$:
\begin{equation}
V(\phi,\beta) = {9 \lambda^2 (\phi^4 - 
   2 \phi^2 (v^2 - \beta^2) + (v^2 + \beta^2)^2)\over 4 \,
(3 - \beta^2)^2} .
\end{equation}
The minimum of this potential with respect to $\beta$ is at $\beta = 0$. 
The curvature of this potential with respect to $\beta$ near $\beta = 0$ is given by
\begin{equation}
V_{\beta,\beta} = {\lambda^2\over 3} (3 (\phi^2+v^2) + (\phi^2 - v^2)^2) >0 .
\end{equation}
During inflation 
\be H^2 =  {\lambda^2\over 12} (\phi^2-v^2)^2
\ee
and therefore
\be
m^2_{\beta} =  4 H^2 +\lambda^2 (v^2 + \phi^2) > 4H^2.
\ee
This means that the field $\beta$ rapidly rolls towards $\beta = 0$ and stays there.

Stability of the $S$ field is not automatic. Its mass squared near the inflationary trajectory $S = 0, \beta = 0$ is given by 
\be
m^2_s = {\lambda^2\over 6} \left((6 \zeta-1) (\phi^2 + v^2)^2 + 12 \phi^2\right) \ .
\ee
During inflation, the leading contribution to the mass is given by the term $(6 \zeta-1) {\lambda^2\over 6}  (\phi^2 + v^2)^2$, which has an absolute value greater than $2\,|6 \zeta-1|\, H^2$. Thus, without the term $\zeta (S\bar S)^2/3$ in the \K\, potential, the effective mass of the field $S$ at large $\phi$ and at $\phi \ll v$ would be equal to $-2H^2$, which would destabilize the inflationary regime. Stability requires that $\zeta > 1/6$.

Moreover, if we want to make the field $S$ heavy enough to avoid inflationary perturbations of the field $S$, we would need to require $6 \zeta-1 \gtrsim 1$, i.e. $\zeta \gtrsim 1/3$.
Under this condition, inflationary evolution occurs along the direction $S = 0$, $\beta = 0$. Since the kinetic term for the field $\phi$ in this direction is canonical, observational consequences of inflation in this model coincide with the observational consequences of inflation in the simple single-field model with the symmetry breaking potential (\ref{inflpot}), see \cite{Kallosh:2007wm,Rehman:2010es}.

There are two possible inflationary regimes in this model. First of all, the field $\phi$ may roll down from  its large values. For  $v \lesssim 1$, inflation is controlled by the term $\lambda^2\phi^4/4$, in which case this model is essentially ruled out by observations, see black circles in Fig. \ref{WMAP}, which shows the predictions of our model \cite{Kallosh:2007wm} as compared to the WMAP results \cite{Komatsu:2010fb}.

However, if $v \gg 1$, the potential during the last stage of inflation is dominated by the quadratic part of the potential near its minimum, with the effective mass squared $2\lambda^2 v^2$. In this case the predictions of the model are the same as of the inflationary theory $m^2\phi^2/2$ with $m = \sqrt 2 \lambda v$, as shown by the white circles in Fig. \ref{WMAP}. The observed amplitude of curvature perturbations of metric requires $\lambda v \sim 2\times 10^{-6}$ \cite{book,Linde:2007fr}.

The upper branch of each blue line in Fig. \ref{WMAP} describes the predictions of the model when the parameter $v$ grows from $v\ll  1$ (black circles) to $v \gg 1$ (white circles). Thus these predictions interpolate between the predictions of the models with a quartic and a quadratic potential.

Inflation may also happen when the field rolls down from $\phi = 0$, as in the new inflationary scenario. This is possible only for $v > 1$. If $v \gg 1$, the stage of inflation describing formation of the observable part of the universe occurs near the minimum of the inflationary potential at $\phi \sim v$, and once again we have the same predictions as in the inflationary theory $m^2\phi^2/2$ with $m = \sqrt 2 \lambda v$, as shown by the white circles in Fig. \ref{WMAP}. On the other hand, if $v$ is not too large, the main part of the inflationary regime occurs near $\phi = 0$, which results in the predictions shown by the lower parts of the blue lines  in Fig. \ref{WMAP}.

There are two blues lines shown in Fig. \ref{WMAP}, corresponding to two different possibilities for the number of e-foldings: $N = 50$ and $N = 60$. In this sense, the predictions of our model, depending on $N$ and $v$, are bounded by these two lines.

\section{Introducing nonminimal coupling to gravity}

The model discussed in the previous section allows various generalizations. For example, one may add to the superpotential a term $\sim S^3$. This will not change the potential along the inflationary trajectory, but it will require a somewhat greater value of $\zeta$ for stabilization of the field $S$. One may also add a term $S\Phi$, which will shift the position of the minimum of the potential along the inflationary trajectory, without altering its shape. 

\begin{figure}[ht!]
\centering
\includegraphics[scale=0.65]{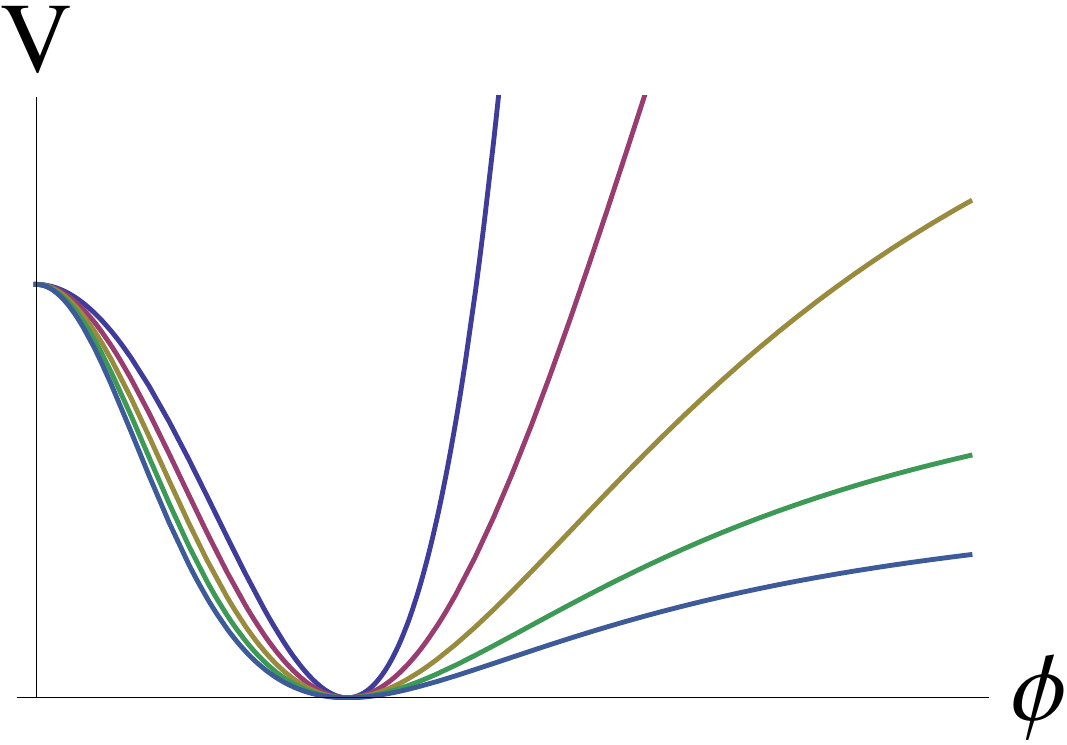}
\caption{Scalar potential (\ref{inflpotmodif}) of the theory (\ref{W}), (\ref{K}) with the nonminimal coupling of the inflaton field ${\xi\over 2} \phi^2 R$. This coupling can be introduced by adding $\xi(\Phi^2+ \bar\Phi^2)$ to $\Omega^2$ and to the \K\, potential. We present a family of potentials for different values of $\xi$, starting from $\xi = 0$ (upper blue line).}
\label{potential2}
\end{figure}

A more interesting change happens if one considers models with $\xi =  -{1/6} +  {\chi/4} > 0$. This is equivalent to adding a term ${\Delta\chi\,\over 4}(\Phi^2+ \bar\Phi^2) = \xi(\Phi^2+ \bar\Phi^2)$ to $\Omega^2$ (\ref{K0}), and, correspondingly, to the \K\, potential (\ref{K1}), under the logarithm. It is also equivalent to making the inflaton field nonminimally coupled to gravity as ${\xi\over 2} \phi^2 R$ \cite{Ferrara:2010yw}. The potential of the scalar field $\phi$ in Einstein frame becomes
\begin{equation}\label{inflpotmodif}
V(\phi) = {\lambda^2\over 4} {(\phi^2-v^2)^2\over (1+\xi \phi^2)^2} ,
\end{equation}
see Eq. (\ref{nc}) and Fig. \ref{potential2}. Note that the field $\phi$ for $\xi \not = 0$ is not canonically normalized. In order to study inflation in this model by standard methods one may perform a transformation to a canonically normalized field $\varphi$ (\ref{canon}), see \cite{Bezrukov:2007ep,Salopek:1988qh,Ferrara:2010in}.

For $\xi \ll 1$, our model will be only slightly modified, but it will allow to further extend the range of the possible values of $n_s$ and $r$. In the limiting case $\xi \gg 1$, for small $v$, the model will have properties similar to the properties of the Higgs inflation models in supergravity discussed in \cite{Einhorn:2009bh,Ferrara:2010yw,Lee:2010hj,Ferrara:2010in}, but for large $v$ this model may allow inflation at small $\phi$ as well.  

An investigation of $n_s$ and $r$ for generic values of $\xi$ in the model for the model ${\lambda\over 4}\phi^4$, which corresponds to our model with $v = 0$, can be found in \cite{Okada:2010jf}. It is interesting that their results are now valid in the context of a new class of supergravity models which we study here. We may therefore use the values of $n_s$ and $r$ which were computed for the case $v=0$ in \cite{Okada:2010jf}. We show these values by green dashed lines in Fig. \ref{WMAP}. Further modifications occur when one takes into account quantum corrections. We expect that by using the freedom of choice of $\xi$ and $v$ in our model one can cover a significant part of the allowed values of $n_s$ and $r$ shown  in Fig. \ref{WMAP}. 

Thus in this class of models the existence of an inflationary regime is rather stable with respect to the change of the parameter $\chi$, or, equivalently, $\xi =  -{1/6} +  {\chi/4}$, but the observational predictions of these models may vary. The model with $\chi = 2/3$, which is equivalent to considering the inflaton field minimally coupled to gravity, $\xi = 0$, is singled out by its simplicity, and by the fact that in this model the \K\, potential vanishes during inflation. In this class of models  the \K\, metric is flat, and the inflaton as well as other scalars have all canonical kinetic terms during inflation. This simple model requires $v \gg 1$ for its consistency with observational data. However, this requirement is no longer needed if one studies generalized versions of this model with sufficiently large values of $\xi$.

\section{Polynomial \K\, potential}\label{quadrK}
It is quite instructive to compare the model considered in the previous section with a similar but different model proposed in \cite{Kawasaki:2001as} (see also closely related models proposed in \cite{Yamaguchi:2000vm,Yamaguchi:2001pw}).
This model has the same superpotential as in the previous section,
\begin{equation}
W=-\lambda S(\Phi^2 -v^2/2).
\label{W2}\end{equation}
The \K\, potential in this model is not logarithmic,
\begin{equation}
\mathcal{K} =  S \bar S - \frac{1}{2}(\Phi -\bar\Phi)^2 - \zeta (S\bar S)^2 .
\label{K2}
\end{equation}
Note that we added to the \K\, potential of the model of \cite{Kawasaki:2001as} the stabilizing term $- \zeta (S\bar S)^2$.

The potential of the inflaton field $\phi = \sqrt 2\, {\rm Re\, \Phi}$ in this theory is the same as in the model studied in the previous section:
\begin{equation}\label{inflpot2}
V(\phi) = {\lambda^2\over 4} (\phi^2-v^2)^2 .
\end{equation}
Ref. \cite{Kawasaki:2001as} studied inflation in this model at small values of the field $\phi$. As we have seen in the previous section, this model incorporates the large field chaotic inflation as well. 

In our model, the mass squared of the imaginary part of the field $\Phi$ (i.e. of the field $\beta$) near the inflationary trajectory with $S = 0$ and $\beta = 0$ is given by
\begin{equation}
m^2_{\beta} =  {\lambda^2\over 2}  ((\phi^2 - v^2)^2+ 2(h^2+v^2)) = 6H^2 + \lambda^2 (\phi^2+ v^2).
\end{equation}
Thus during inflation $m^2_{\beta} > 6H^2$, and therefore the imaginary part of the field $\Phi$ is indeed stabilized at ${\rm Im}\, \Phi = 0$, i.e. at $\beta = 0$.

The mass of the field $S$ along the inflationary trajectory $S = 0$ is 
\begin{equation}
m^2_{s} =  12 \zeta H^2  + 2\lambda^2 \phi^2.
\end{equation}
Therefore the classical part of the field $S$ is also stabilized at $S = 0$. However, for $\zeta = 0$, as in the original model of Ref. \cite{Kawasaki:2001as}, the mass of the field $S$ during inflation is much smaller than the Hubble constant, which means that the field $S$ must experience large inflationary fluctuations \cite{Davis:2008fv}. Depending on the coupling of the field $S$ to matter, these fluctuations may lead to dangerous isocurvature perturbations of the metric, or to non-gaussian adiabatic perturbations, as in the theory of the curvaton field \cite{curva}.  The amplitude of these perturbations and the degree of non-gaussianity may vary, depending on the choice of the parameter $\zeta$ and on the initial conditions for inflation  \cite{Demozzi}. 

Stabilization of the field $S$, which is achieved if $\zeta \gtrsim 0.1$, provides a significant simplification of the inflationary theory. Indeed, by adding the term $-\zeta (S\bar S)^2$ to the \K\, potential we can render the mass of the field $S$ greater than $H$, which eliminates inflationary perturbations of this field and reduces the problem to the investigation of single-field inflationary cosmology for a potential ${\lambda^2\over 4} (\phi^2-v^2)^2$. Observational consequences of this model, after the stabilization of the field $S$, coincide with the observational consequences of the model studied in the previous section.

The properties of this model change dramatically if one modifies it by adding to its \K\, potential the term $\xi(\Phi^2+ \bar\Phi^2)$, as we did in the end of the previous section. This term introduces the overall factor $e^{-3\xi\phi^2}$ into the inflaton potential, see Eq. (\ref{ncpol}). If we want to preserve the previously developed scenario and use it for the description of the last 60 e-folds of inflation, we must have $|3\xi\phi^2| \lesssim 1$ for $|\phi | \lesssim 10$. This leads to the constraint $|\xi| \lesssim 3\times 10^{-3}$, which is a significant fine tuning. This problem was absent in the model studied in the previous section.

In this sense, the new class of theories with the logarithmic \K\, potential  discussed in Sect. \ref{log} allows a much greater flexibility in the construction of models of chaotic inflation. On the other hand, as we already mentioned, the model considered in the present section may allow us to obtain non-gaussian adiabatic perturbations related to the field $S$. The amplitude and the degree of non-gaussianity in this model can be controlled by the choice of the stabilization parameter $\zeta$.

\section{Chaotic inflation with a quadratic potential}\label{logK}

Now we will consider a very simple version of the chaotic inflation model with the superpotential
\begin{equation}
W=m S\Phi,
\label{W3}\end{equation}
and the polynomial \K\, potential (\ref{K2})
\begin{equation}
\mathcal{K} =  S \bar S - \frac{1}{2}(\Phi -\bar\Phi)^2 - \zeta (S\bar S)^2 .
\label{K3}
\end{equation}
This model almost coincides with the model of chaotic inflation in supergravity proposed in \cite{Kawasaki:2000yn}. The main difference is the presence of the stabilizing term $- \zeta (S\bar S)^2$ in the \K\, potential of our model. Also, to bring this model closer to the notations of \cite{Einhorn:2009bh,Ferrara:2010yw,Lee:2010hj,Ferrara:2010in}, we use the term  $- \frac{1}{2}(\Phi -\bar\Phi)^2$ instead of the term $ \frac{1}{2}(\Phi +\bar\Phi)^2$ in the \K\, potential of the model of Ref. \cite{Kawasaki:2000yn}. As a result, the role of the inflaton  is played in our scenario by the real part $\phi$ of the field $\Phi$, instead of its imaginary part $\beta$. 

In terms of the variables $S=(s+i\alpha)/\sqrt 2$, $ \Phi=(\phi+i\beta)/\sqrt 2$, the potential grows as $e^{|S|^2}$ at large $|S|$, and as $e^{\beta^2}$ at large $\beta$. However, along the inflationary trajectory at $S= \beta = 0$, the potential of the inflaton field $\phi$ in this theory is quadratic \cite{Kawasaki:2000yn}, as in the simplest version of the chaotic inflation scenario \cite{Linde:1983gd}:
\begin{equation}\label{inflpot3}
V(\phi) = {m^2\over 2} \phi^2.
\end{equation}
The proper amplitude of perturbations of metric requires $m \sim 2 \times 10^{-6}$ \cite{book,Linde:2007fr}.

Near the inflationary trajectory with $S = 0$, the mass squared of the imaginary part of the field $\Phi$, or, equivalently, of the field $\beta$, is 
\begin{equation}
m^2_{\beta} =  m^2(1 + \phi^2) = 6H^2 + m^2.
\end{equation}
Thus during inflation the imaginary part of the field $\Phi$ is stabilized at ${\rm Im}\, \Phi = 0$, or equivalently, $\beta = 0$.

The mass of the field $S$ along the inflationary trajectory $S = 0$ is given by
\begin{equation}
m^2_{s} =  12 \zeta H^2  + m^2 .
\end{equation}
For $\zeta = 0$, the mass squared of the field $S$ is equal to $m^2$, which is much smaller than $H^2$ during inflation.  This means that the field $S$ (both its real and imaginary part) must experience large inflationary fluctuations \cite{Davis:2008fv}. Depending on the coupling of the field $S$ to matter, these fluctuations may lead to dangerous isocurvature perturbations of metric, or to non-gaussian adiabatic perturbations, as in the curvaton theory \cite{curva}.  By adding the term $-\zeta (S\bar S)^2$ to the \K\, potential of this model one can render the mass of the field $S$ greater than $H$. This eliminates inflationary perturbations of the field $S$ and reduces the problem to the investigation of single-field inflationary cosmology for a potential $ {m^2\phi^2\over 2}$ \cite{Linde:1983gd}.

This model, just as the model considered in the previous section, is not very flexible with respect to the addition of the term ${\xi}(\Phi^2+ \bar\Phi^2)$ to the \K\, potential (\ref{K3}). This term significantly changes the potential at large $\phi$:
\be\label{ncpol22}
V(\phi) = {m^2\over 2} \phi^2\  e^{-3\xi\phi^2}  \ .
\ee 
This model can successfully describe the last 60 e-folds of inflation only if the exponential term does not significantly alter the potential for $\phi \lesssim 15$, which implies that one should take $\xi \lesssim 10^{-3}$.

One can easily generalize this model and incorporate it into the class of the models considered in  
\cite{Einhorn:2009bh,Ferrara:2010yw,Lee:2010hj,Ferrara:2010in}. We will use the same superpotential $W=m S\Phi$ as in Eq. (\ref{W3}), but the logarithmic  \K\, potential (\ref{K}),
\begin{equation}
\mathcal{K} = -3\log \left[ 1 + \frac{1}{6}(\Phi -\bar\Phi)^2 - \frac{1}{3}S \bar S+ \zeta (S\bar S)^2/3\right].
\label{K3a}
\end{equation}

This theory has the same inflaton potential $m^2\phi^2/2$. 
Near the inflationary trajectory with $S = 0$, the mass squared of the imaginary part of the field $\Phi$, or, equivalently,  is 
\begin{equation}
m^2_{\beta} =  m^2(1 + {2\over 3}\phi^2) = 4H^2 + m^2.
\end{equation}

The mass squared of the field $s$ along the inflationary trajectory $S = 0$ is 
\begin{equation}
m^2_{s} =  2 (6\zeta-1) H^2  + m^2 .
\end{equation}
Thus we will have a successful inflationary model if $\zeta$ is somewhat greater than $1/6$.

\begin{figure}[ht!]
\centering
\includegraphics[scale=0.65]{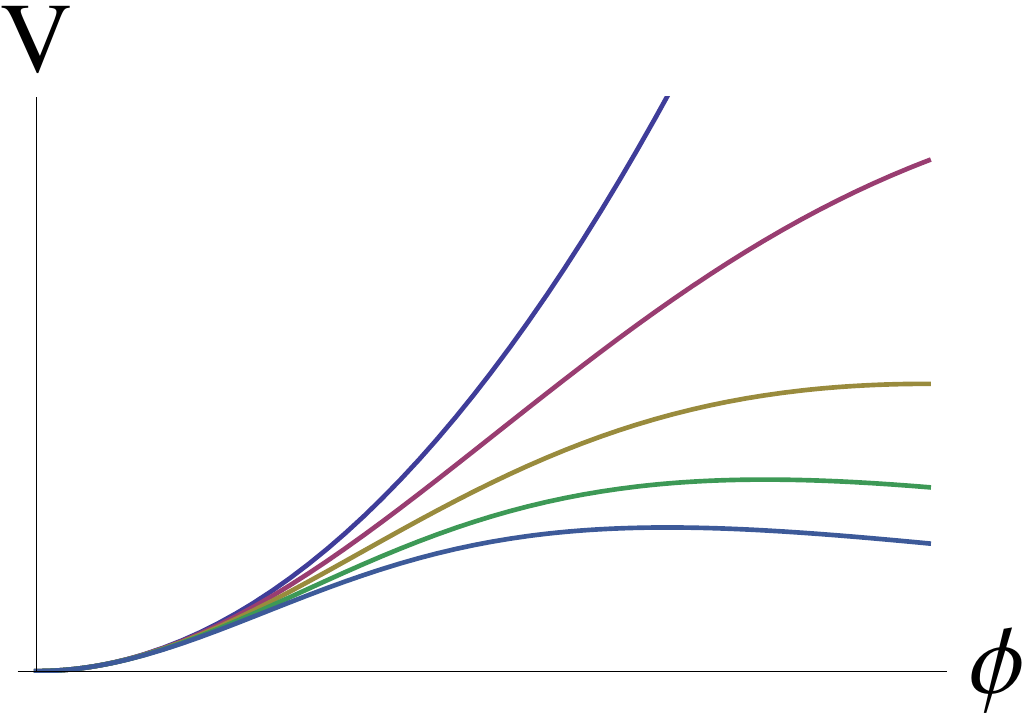}
\caption{Scalar potential (\ref{inflpot3a}) of the theory (\ref{W3}), (\ref{K3a}) with the nonminimal coupling of the inflaton field ${\xi\over 2} \phi^2 R$.  We present a family of potentials for different values of $\xi$, starting from $\xi = 0$ (upper blue line).}
\label{potentialfig2}
\end{figure}

One can introduce nonminimal coupling of the inflaton field by adding the term ${\xi}(\Phi^2+ \bar\Phi^2)$ under the logarithm in the expression for the \K\, potential (\ref{K3a}). In this case, the Einstein frame potential becomes
\begin{equation}\label{inflpot3a}
V(\phi) = {m^2\phi^2\over 2(1+\xi \phi^2)^2} ,
\end{equation}
see Fig. \ref{potentialfig2}. For $\xi \ll 1$, this potential has a flat maximum at $\phi = 1/\sqrt \xi \gg 1$. Depending on the value of $\xi$, the last 60 e-foldings of inflation in this scenario occur either near the maximum of the potential  at $\phi = 1/\sqrt \xi$, or during slow roll near the quadratic part of the potential, at $\phi <  1/\sqrt \xi$. An interpolation between these two regimes, which is achieved by changing of the small parameter $\xi$, produces a continuous variety of values of $n_s$ and $r$, similar to the variety shown by the blue lines in Fig. \ref{WMAP}. To make this model phenomenologically viable one should have $|\xi| \lesssim 10^{-2}$.

\section{Functional freedom of choice of the inflaton potential}\label{general}

Now we will consider a more general superpotential
\begin{equation}
W= S f(\Phi),
\label{W3gen}\end{equation}
and the \K\, potential
\begin{equation}
\mathcal{K} =  S \bar S - \frac{1}{2}(\Phi -\bar\Phi)^2 - \zeta (S\bar S)^2 .
\label{K3ss}
\end{equation}
As we already shown in Sect. {\ref{CSS}}, the potential of the inflaton field $\phi$ in this theory, for $S = 0$ and ${\rm Im}\, \Phi = 0$, is given by
\begin{equation}\label{inflpot3general}
V(\phi) = |f({\phi/ \sqrt 2})|^{2}.
\end{equation}

A particular example is given by superpotentials which at large $\Phi$ look as 
\begin{equation}
W =  \lambda_{\alpha}\, S\, \Phi^{\alpha} ,
\end{equation}
which lead to inflationary potentials\footnote{This possibility was pointed out to us by A. Westphal.}
\begin{equation}\label{simplepower}
V =  {\lambda^{2}_{\alpha}\over 2^{\alpha}} \,  \phi^{2\alpha} .
\end{equation}
For $V  \sim \phi^{2\alpha}$, one has $1-n_{s} =(1+\alpha) N^{-1}$, $r = 8\alpha/N$, where $N$ is the number of e-folds.
 In this case $1-n_{s} = 1/N + r/8$  \cite{Silverstein:2008sg,Alabidi:2010sf}.
 
  \begin{figure}[t!]
\centering
\includegraphics[scale=0.32]{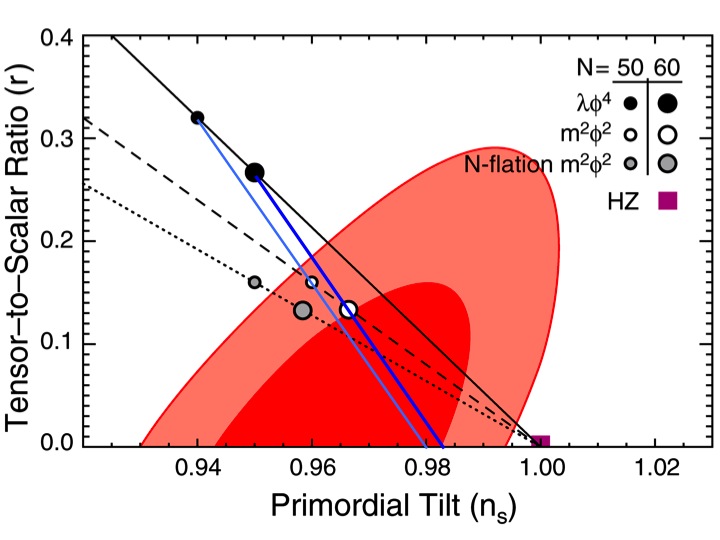}
\caption{Predictions of the large field inflation in the model with the potential $V\sim (\phi^\alpha-v^\alpha)^2$ for $v < 1$ and different $\alpha$ in the range of $0< \alpha <2 $ are shown by two blue lines corresponding to $N = 60$ and $N = 50$.}
\label{WMAP2}
\end{figure}

 The problem with this class of potentials is that for $0<\alpha < 1$ the mass of the field $\phi$ is singular at the minimum of the potential. Meanwhile for $\alpha < 0$ the potential gradually decreases at $\phi \to \infty$, so there is no stable vacuum state. One can avoid these problems in the generalized version of the model (\ref{W}), with the superpotential
 \begin{equation}
W=-\lambda_\alpha S\left(\Phi^\alpha -{v^{\alpha}\over 2^{\alpha/2}}\right) 
\label{Wa}\end{equation}
and the inflaton potential
\begin{equation}\label{inflpot2aa}
V(\phi) = {\lambda_\alpha^2\over 2^{\alpha}} (\phi^\alpha-v^\alpha)^2 .
\end{equation}
This potential has a non-singular minimum at $\phi = v$. If $v$ is not too large, at $\alpha > 0$ one has the same regime of chaotic inflation as in the theory ${\lambda^{2}_{\alpha}\over 2^{\alpha}} \,  \phi^{2\alpha}$ at large $\phi$. Therefore in this regime the predictions of our model are the same as of the model $ \phi^{2\alpha}$. On the WMAP graph, the family of all possible values of $n_{s}$ and $r$ for this class of models is shown by the two straight lines (for $N = 60$ and for $N = 50$) crossing the points corresponding to the theories ${\lambda\over 4}\phi^{4}$ and ${m^{2}\over 2}\phi^{2}$, see Fig. \ref{WMAP2}.

On the other hand, for $\alpha < 0$ the potential at large $\phi$ is very flat, asymptotically approaching $ {\lambda_\alpha^2\over 2^{\alpha}}\, v^{2\alpha}$,
\begin{equation}\label{inflpotbb}
V(\phi) = {\lambda_\alpha^2\over 2^{\alpha}} \left(v^{2\alpha}\left(1 - 2\left({\phi/ v}\right)^{-|\alpha|}\right)\right) +...  .
\end{equation}
One can have inflation at large field $\phi$ in this model. Depending on $v$ and $\alpha$, one may also have inflation at $\phi < v$. 

Note that the same potentials (\ref{simplepower}), (\ref{inflpot2}) appear in the theories with the logarithmic \K\, potential (\ref{K}) as well. One can further generalize such models by  making the field $\phi$ nonminimally coupled to gravity. In this case the potential (\ref{inflpot2}) transforms into
\begin{equation}\label{inflpot2bb}
V(\phi) = {\lambda_\alpha^2\over 2^{\alpha}} {(\phi^\alpha-v^\alpha)^2\over  (1+\xi \phi^2)^2}.
\end{equation}
A more detailed discussion of the models (\ref{Wa})-(\ref{inflpot2bb}) will be contained in a separate publication.

More generally, eq. (\ref{inflpot3general}) implies that one may take any holomorphic function $f(\Phi)$ and check whether inflation is possible in the theory with the potential $ |f({\phi/ \sqrt 2})|^{2}$. If this is the case, and if the full supergravity potential $V(\Phi, S)$ has a stable direction with $S = 0$ and ${\rm Im}\, \Phi = 0$, then one may implement chaotic inflation scenario in this theory, with the inflaton potential $V(\phi) =  |f({\phi/ \sqrt 2})|^{2}$.

We begin with the general investigation of stability with respect to the field $S$. The mass squared of the field $S$ along the inflationary trajectory $S = 0$, ${\rm Im}\, \Phi = 0$ is given by
\begin{equation}
m^2_{s} =  12 \zeta H^2  + (f'(\phi/\sqrt 2))^{2} .
\end{equation}
Therefore $m^2_{s} > 0$ and the field $S$ is stable for $\zeta \geq 0$.
For $\zeta > 1/12$, one has $m^2_{s} > H^{2}$ in  this class of models, so the perturbations of the field $S$ are not excited during inflation.

The next condition to check is the stability with respect to the imaginary part of the field $\Phi$. Near the inflationary trajectory with $S = 0$, the mass squared of the imaginary part of the field $\Phi$ is 
\begin{equation}
m^2_{\beta} =   {2}(f(\phi/\sqrt 2))^{2} + (f'(\phi/\sqrt 2))^{2} - f(\phi/\sqrt 2)\,  f''(\phi/\sqrt 2) ,
\end{equation}
where the derivatives are taken with respect to ${\rm Re}\, \Phi = \phi/\sqrt 2$. For any function $f$ monotonously growing at large $\phi$, the leading contribution is given by the first term, which is equal to $6H^{2}$. Therefore this stability condition is automatically satisfied for the large-field chaotic inflation. 

In general, one should also check whether the stability condition is satisfied at $\phi \lesssim 1$, when the large-field inflation ends. This may lead to certain constraints on the choice of the function $f(\phi)$ at small $\phi$, but this should not affect the freedom of choice of $f(\phi)$ during the large field chaotic inflation. Moreover, the requirement $m^2_{\beta} > 0$ is not absolutely necessary after the end of inflation. First of all, the post-inflationary evolution may never bring the field into the region where $m^2_{\beta} < 0$. Moreover, at $S=0$, the potential of the field $\Phi$ becomes
\begin{equation}
V(\Phi) =   e^{2 ({\rm Im}\, \Phi)^2} |f(\Phi)|^2 .
\end{equation}
Therefore the point where the function $f(\phi)$ vanishes at some real value of $\Phi$ corresponds to the absolute minimum of the potential along the trajectory $S = 0$, so the fields eventually tend to move towards this minimum. In general, it might happen that at small $\phi$ the model is unstable with respect to generation of the field $\beta$, which triggers an instability of the field $S$. This does not seem to be a generic possibility, but one should check it nevertheless for each particular inflationary model.

One can generalize this scenario and incorporate it in the class of the models considered in  
\cite{Einhorn:2009bh,Ferrara:2010yw,Lee:2010hj,Ferrara:2010in}. One may use the same superpotential as before, 
\begin{equation}
W= S f(\Phi),
\label{W3gena}\end{equation}
 but instead of the power-law \K\, potential we will use the  logarithmic  \K\, potential (\ref{K}),
\begin{equation}
\mathcal{K} = -3\log \left[ 1 + \frac{1}{6}(\Phi -\bar\Phi)^2 - \frac{1}{3}S \bar S+ \zeta (S\bar S)^2/3\right].
\label{K3aa}
\end{equation}
This theory has the same inflaton potential $|f({\phi/ \sqrt 2})|^{2}$. 

The mass squared of the field $S$ along the inflationary trajectory $S = 0$, ${\rm Im}\, \Phi = 0$ is 
\begin{equation}
m^2_{s} =  2 (6\zeta-1) H^2  +  (f'(\phi/\sqrt 2))^{2} .
\end{equation}
This means that the regime $S = 0$ is stable for $\zeta > 1/6$.

The mass squared of the field $\beta$ is 
\begin{equation}
m^2_{\beta} =   {4\over 3}(f(\phi/\sqrt 2))^{2} + (f'(\phi/\sqrt 2))^{2} - f(\phi/\sqrt 2)\,  f''(\phi/\sqrt 2).
\end{equation}
 For $\phi \gg 1$, the first term gives the leading contribution, which is equal to $4H^{2}$. Therefore this stability condition is satisfied for the large-field chaotic inflation. As for the small values of the field $\phi$, one should note that for $S=0$ the potential becomes
 \begin{equation}
V(\Phi) =   {|f(\Phi)|^2\over (1-\beta^2/3)^2} .
\end{equation}
Therefore, just as in the case with the power-law \K\, potential, the point where $f(\phi) = 0$ will correspond to the absolute minimum of the potential along the stable direction $S = 0$. 

Thus, under certain conditions it is possible to have a stable inflationary trajectory with $S = 0$, ${\rm Im}\, \Phi = 0$ for a broad class of superpotentials (\ref{W3gen}). This means that there is a functional freedom of choice of the chaotic inflation potential in the theories discussed in our paper. And one may still further generalize these theories by introducing the nonminimal coupling of the inflaton field to gravity using the \K\, potential (\ref{AL}), in which case the inflaton potential will have the general structure represented in Eq. (\ref{nc}).

\section{Conclusions}
In this paper we continued the investigation of a class of supergravity models proposed to implement  Higgs field inflation, where the Higgs field is non-minimally coupled to gravity \cite{Einhorn:2009bh,Ferrara:2010yw,Lee:2010hj,Ferrara:2010in}. We found that in this class of models the scalar potential has flat directions protected by shift symmetries of the \K\, potential if the inflaton field minimally couples to gravity. We found that the mechanism of stabilization of the inflationary trajectory developed in \cite{Lee:2010hj,Ferrara:2010in} can be applied to these models, and it can be used to improve several other models of chaotic inflation in supergravity. 

It is interesting to compare  different models of chaotic inflation. If one considers the new class of models  with the logarithmic \K\, potential and the parameter $\chi \not = 2/3$, this will affect the flatness of the \K\, potential in the direction of $\Phi +\bar\Phi$, but this change may even improve the model studied in section \ref{log} and bring it, in the large $\chi$ limit, to the form studied in  \cite{Einhorn:2009bh,Ferrara:2010yw,Lee:2010hj,Ferrara:2010in}. If the change of $\chi$ is not too large, it will slightly modify the model discussed in section \ref{logK}. In particular, it will change the observable quantities $n_s$ and $r$.  Meanwhile the models with the power-law \K\, potential are much more sensitive with respect to modifications of the \K\, potential. The terms affecting flat directions of the power-law \K\, potential will cause the scalar potential to grow exponentially at large $\phi$, which will make chaotic inflation more difficult to achieve. In this sense, chaotic inflation in the new models with the logarithmic \K\, potential is  more robust with respect to modifications of the models.

To conclude, we proposed a  new broad class of models of chaotic inflation in supergravity. In these models one has a functional freedom to modify the inflaton potential and, consequently, the values of $n_s$ and $r$. In addition, one can control the degree of non-gaussianity of perturbations of the field $S$, as well as the amplitude of these perturbations, by changing the parameter $\zeta$.  This flexibility may be very important for an interpretation of new and upcoming observational data. 

\

\noindent{\large{\bf Acknowledgments}}

\noindent  We are grateful to S.~Ferrara, S.~Kachru, A.~Marrani, V.~Mukhanov, A.~Van Proeyen, S.~Shenker, E.~Silverstein,  L.~Susskind and A.~Westphal for many enlightening discussions.   This work was supported by NSF grant PHY-0756174.

\end{document}